\documentclass[twocolumn,showpacs,superscriptaddress,preprintnumbers,amsmath,amssymb,prb,floatfix]{revtex4}

\usepackage{graphicx}% Include figure files
\usepackage{dcolumn}% Align table columns on decimal point
\usepackage{bm}% bold math
\usepackage{amsmath}
\usepackage{amssymb}
\usepackage{booktabs}
\usepackage{color}

\begin{document}

\preprint{APS/123-QED}

\title{\textit{Ab initio} study of low-energy electronic collective excitations in bulk Pb}

\author{X. Zubizarreta}
   \affiliation{Donostia International Physics Center (DIPC), Paseo de Manuel Lardizabal 4, 20018
   San Sebasti\'an/Donostia, Basque Country, Spain}
   \affiliation{Departamento de F\'{\i}sica de Materiales, Facultad de Ciencias Qu\'{\i}micas,
   Universidad del Pa\'{\i}s Vasco/Euskal Herriko Unibertsitatea, Apdo. 1072, 20080 San Sebasti\'an/Donostia,
   Basque Country, Spain}
\author{V. M. Silkin}
   \affiliation{Donostia International Physics Center (DIPC), Paseo de Manuel Lardizabal 4, 20018
   San Sebasti\'an/Donostia, Basque Country, Spain}
   \affiliation{Departamento de F\'{\i}sica de Materiales, Facultad de Ciencias Qu\'{\i}micas,
   Universidad del Pa\'{\i}s Vasco/Euskal Herriko Unibertsitatea, Apdo. 1072, 20080 San Sebasti\'an/Donostia, Basque Country,
Spain}
   \affiliation{IKERBASQUE, Basque Foundation for Science, 48011 Bilbao, Spain}
\author{E. V. Chulkov}
   \affiliation{Donostia International Physics Center (DIPC), Paseo de Manuel Lardizabal 4, 20018
   San Sebasti\'an/Donostia, Basque Country, Spain}
   \affiliation{Departamento de F\'{\i}sica de Materiales, Facultad de Ciencias Qu\'{\i}micas,
   Universidad del Pa\'{\i}s Vasco/Euskal Herriko Unibertsitatea, Apdo. 1072, 20080 San Sebasti\'an/Donostia,
   Basque Country, Spain}
   \affiliation{Centro de F\'{\i}sica de Materiales CFM - Materials Physics Center MPC, Centro Mixto CSIC-UPV/EHU,
   Paseo de Manuel Lardizabal 5, 20018 San Sebasti\'an/Donostia, Basque Country, Spain}
\date{\today}

\begin{abstract}
A theoretical study of the dynamical dielectric response of bulk lead at low energies is presented. The calculations are performed with full inclusion of the electronic band structure calculated by means of a first-principles pseudopotential approach. The effect of inclusion of the spin-orbit  interaction in the band structure on the excitation spectra in Pb is analyzed, together with dynamical exchange-correlation and local-field effects. In general, results show significant anisotropy effects on the dielectric response of
bulk Pb. At small momentum transfers in all three symmetry
directions the calculated excitation spectra present several peaks
with strong acoustic-like dispersion in the energy range below 2 eV.
The analysis shows that only one of such modes existing at momentum
transfers along the $\Gamma$-K direction can be interpreted as a
true acoustic-like plasmon whereas all other modes are originated
from the enhanced number of intraband electron-hole excitations at
corresponding energies. Comparison with available optical
experimental data shows good agreement.
\end{abstract}


\maketitle

\section{INTRODUCTION}

In the last two decades significant progress was achieved in the
understanding of
dynamical processes involving both single particles -- electrons and holes --
and collective electronic excitations in metallic systems from the so-called
{\it ab initio} perspective.
In particular, in the course of the study of dynamics of electronic
excitations the relative importance of such intrinsic processes
as electron-phonon and electron-electron decay was thoroughly
analyzed.\cite{ChulkovCR06,ecbessr04} One of the important results of these investigations was the demonstration of
a crucial role of electronic band structure of real solids in such events.

At the same time, it is known that the band structure can be
influenced by spin-orbit (SO) effects, especially in heavy elements. Remarkable examples
are lead, bismuth and bismuth tellurohalides, where the SO interaction produces strong modifications of the
band structure \cite{gonzebi,zubizarreta,Eremeev} and vibrational spectra,\cite{heboprb10,biph,Sklyadneva} influencing the
electron-phonon interaction in these materials.

The effect of the SO interaction on the electronic structure was
intensively investigated in the past, while the inclusion of this
interaction into calculation of dielectric properties from
first-principles was performed in few cases only. Thus it was demonstrated that the
SO interaction induces sizeable effects in the optical properties and dielectric properties in the small momentum transfer
limit of heavy elements, in particular Pb.\cite{glamnjp10}
However, a detailed study of the impact of the SO interaction on the
excitation spectrum of Pb over a whole momentum-energy domain is
still missing. Especially in a low-energy region where the major effect is
expected. This constitutes a main topic of the present study, in such a way giving a
whole picture of the low-energy electronic dynamics in the elemental lead in addition to the
phonon dynamics studied heavily up to now.

One of the characteristics of lead is that it presents the
second-highest critical temperature of all bulk elemental
superconductors ($T_{c}$ = 7.23 K),\cite{dyroprb75} which has been shown to be
related to its large electron-phonon coupling (EPC) constant. In
addition, SO coupling has a strong impact on the
electron-phonon interaction in bulk Pb, increasing its strength as
much as $44\%$.\cite{heboprb10} Studying the electronic collective
excitations near the Fermi level ($E_F$) and the SO effects on them
would complete the description of the low-energy dynamics in bulk lead,
and it is done in the present work.

For many years the electron-density response of solids was studied
using a free-electron gas (FEG) model in which the electron valence
density is parameterized by a single quantity: the density parameter
$r_s$, which stands for the average inter-electron
distance.\cite{pino66} The FEG model gave insight into basic
properties of the momentum- and frequency-dependent dynamical
dielectric response. However, band structure effects that are
missing in a FEG model frequently can produce strong impact on the
dynamical dielectric response of solids. In particular, interband
transitions (not presented in a FEG model) give rise, for instance,
to a strong red shift of the Ag plasmon frequency \cite{cazaliprb00}
or to a negative momentum dispersion of the plasmon in
Cs.\cite{aryaseprl94} Moreover, these transitions can dominate in some materials the
energy-loss landscape in the low-energy-transfer domain.\cite{zhsiprb01,kupiprl02,ecchprb12}
%Also, it
%has been shown that band structure effects play an important role in
%the screening of charges at noble metal surfaces.\cite{diezpnas11}

In three-dimensional (3D) solids the FEG model predicts the
existence of a $r_s$-dependent threshold for collective excitations,
exceeding in metals several eV. Hence, according to the FEG theory,
plasmons can not participate directly in the low-energy dynamical
processes near the Fermi surface. However, in the fifties it was
predicted\cite{pines56,pinespr58} the existence of a very-low energy
excitation, which should be present in systems with several energy
bands crossing the Fermi level with different Fermi velocities,
$v_F$, as it is the case of bulk Pb. This very-low energy mode
presents an acoustic-like dispersion at small momentum transfers $q$'s, i.e.,
$\omega_{\rm AP}=v_{\rm AP}\cdot q$, where $v_{\rm AP}$ is the group
velocity of the acoustic plasmon being very close to the Fermi
velocity in the energy band with the slower carriers. Thus, the
$\omega_{\rm AP}$ frequency tends to zero as $q\rightarrow0$.
Exchange of acoustic plasmons have been suggested as a possible
mechanism of electron pairing in superconductors (see, i.e., Ref.
\onlinecite{ishiiprb93} and references therein). Nevertheless, to
the best of our knowledge this kind of acoustic-like modes has not
been demonstrated to exist in bulk metallic systems experimentally.
Similar mode has been recently predicted \cite{euro,Silkin05} and proved experimentally\cite{nature,papaprl10,podiepl10} on metal surfaces.
On the other hand, recent detailed \textit{ab initio} calculations of the dynamical dielectric response in a variety of bulk metallic systems like MgB$_2$,\cite{silkinprb09a,balassprb08} Pd,\cite{silkinprb09b} transition-metal dichalcogenides,\cite{faarprb12,cugaprb12} and CaC$_6$\cite{ecchprb12} predicted, with some initial controversy,\cite{zhsiprb01,kupiprl02} the existence of a such kind of acoustic plasmons in metallic systems. In particular, in the case of Pd and CaC$_6$ this acoustic-like mode disperses in all three symmetry directions while in the layered compounds MgB$_2$ and NbSe$_2$ the corresponding mode exists only for the momentum transfers along the direction perpendicular to the basal planes.

%Recently the dielectric properties at small momentum transfers and
%optical properties of Pb were investigated in detail theoretically
%by {\it ab initio} calculations and experimentally by reflection
%electron energy-loss spectroscopy
%(REELS).\cite{wegljpcrd09,glamnjp10} Therefore it would be also
%useful to compare those data with the ones obtained by other
%calculation method.

The main aim of the present study is twofold. First, the complexity
of the low-energy electronic response of bulk Pb is demonstrated, together
with the role of different physical ingredients. Second, the
results on the modes characterized by an acoustic-like dispersion are presented and analyzed in detail.
In particular we demonstrate strongly anisotropic character of
such low-energy collective excitations in bulk lead. The
calculations of the low-energy collective excitations have been done
using the first-principles pseudopotential approach within the
time-dependent density functional theory
(TDDFT).\cite{rungeprl84,petersprl96}

The rest of the paper is organized as follows: In Sec. II the
details of the \textit{ab initio} calculations of the dynamical
dielectric response of 3D solids are presented. In Sec. III the
general results for bulk Pb for the low-energy regime are presented together
with comparison with experimental optical data, while in Sec. IV
the acoustic-like collective excitations are analyzed in detail. Finally, the
main conclusions are presented in Sec. V. Unless otherwise stated,
atomic units are used throughout, i.e., $e^2=\hslash=m_e=1$.

\section{CALCULATION METHOD}

A key quantity in the description of dynamical dielectric response properties
of solids is the dynamical structure factor $S(\textbf{Q},\omega)$
since, within the first Born approximation, the inelastic scattering
cross section at momentum transfer ${\bf Q}$ and energy $\omega$ of X-rays and electrons is proportional to it.\cite{pino66}
$S(\textbf{Q},\omega)$ is related by the fluctuation-dissipation
theorem to the dielectric function
$\varepsilon(\textbf{r},\textbf{r}',\omega)$. For a periodic solid,
\begin{equation} \label{eq:eels}
S(\textbf{Q},\omega) = -\frac{\Omega
|\textbf{q}+\textbf{G}|^2}{2\pi}{\rm
Im}[\varepsilon^{-1}_{\textbf{G},\textbf{G}}(\textbf{q},\omega)],
\end{equation}
where $\Omega$ is the normalization volume, \textbf{G} is a
reciprocal-lattice vector, vector ${\bf q}$ is located in the
Brillouin zone (BZ), $\textbf{Q}=\textbf{q}+\textbf{G}$, and ${\rm
Im}[\varepsilon^{-1}_{\textbf{G},\textbf{G}'}(\textbf{q},\omega)]$
is the so-called energy-loss function, whose Fourier coefficients are
related to those of the density-response function for interacting
electrons $\chi(\textbf{r},\textbf{r}',\omega)$ through
\begin{equation} \label{eps_inv}
\varepsilon^{-1}_{\textbf{G},\textbf{G}'}(\textbf{q},\omega) =
\delta_{\textbf{G},\textbf{G}'}
 + v_{\textbf{G}}(\textbf{q})\chi_{\textbf{G},\textbf{G}'}(\textbf{q},\omega),
\end{equation}
where $v_{\textbf{G}}(\textbf{q}) =
\dfrac{4\pi}{|\textbf{q}+\textbf{G}|^2}$ is the Fourier transform of
the bare Coulomb interaction. Note that in this definition the ${\bf
Q}$ vector is not restricted to be located inside the first BZ.

A crucial quantity in the evaluation of Eq. (\ref{eps_inv}) is the
density-response function $\chi$, which in the  framework of TDDFT
\cite{rungeprl84,petersprl96} satisfies the matrix equation
\begin{multline} \label{chi}
\chi_{{\bf G},{\bf G}'}({\bf q},\omega) = \chi^{\rm o}_{{\bf G},{\bf G}'}({\bf q},\omega) + \sum_{{\bf
G}''}\sum_{{\bf G}'''} \chi^{\rm o}_{{\bf G},{\bf G}''}({\bf q},\omega) \\
\times [v_{{\bf G}''}({\bf q})\delta_{{\bf G}'',{\bf G}'''} + K^{\rm
XC}_{{\bf G}'',{\bf G}'''}({\bf q})] \chi_{{\bf G}''',{\bf G}'}({\bf
q},\omega),
\end{multline}\\
where $\chi^{\rm o}_{{\bf G},{\bf G}'}({\bf q},\omega)$ is the
matrix of the Fourier coefficients of the  density-response function
for noninteracting Kohn-Sham electrons. $K^{\rm XC}_{{\bf G},{\bf
G}'}({\bf q})$ stands for the Fourier components of the
exchange-correlation kernel, whose exact form is unknown. Thus,
approximations must be used to describe $K^{\rm XC}_{{\bf G},{\bf
G}'}({\bf q})$.  Here we use two frequently used approaches for the
description of $K^{\rm XC}$: the random-phase approximation (RPA)
(here one simply sets $K^{\rm XC}_{RPA}$ to zero, i.e., neglects the
short-range exchange and correlation effects) and the time-dependent
local-density approximation (TDLDA).\cite{gross96}

We evaluate the $\chi^0_{\textbf{G},\textbf{G}'}(\textbf{q},\omega)$
matrix by calculating first the  spectral function matrix
$S^0_{\textbf{G},\textbf{G}'}(\textbf{q},\omega)$ using the following
expression\cite{aryaseprb94,aryase01}
\begin{widetext}
\begin{equation} \label{str_fac}
S^0_{\textbf{G},\textbf{G}'}(\textbf{q},\omega)=
\frac{1}{\Omega}\sum_{\textbf{k}}^{\rm BZ}\sum_{n}^{\rm
occ}\sum_{n'}^{\rm unocc}\langle\psi_{n\textbf{k}}|e^{-{\rm
i}(\textbf{q}+\textbf{G})
\cdot\textbf{r}}|\psi_{n'\textbf{k}+\textbf{q}}\rangle\langle\psi_{n'\textbf{k}+\textbf{q}}|e^{{\rm
i}(\textbf{q}+\textbf{G}')
\cdot\textbf{r}}|\psi_{n\textbf{k}}\rangle\delta(\varepsilon_{n\textbf{k}}-\varepsilon_{n'\textbf{k}+\textbf{q}}+\omega).
\end{equation}
\end{widetext}
In Eq. (\ref{str_fac}), $n$ and $n'$ are band indexes, wave vectors
\textbf{k}'s are in the first BZ, and $\varepsilon_{n\textbf{k}}$
and $\psi_{n\textbf{k}}$ are Bloch eigenvalues and eigenfunctions,
respectively, of the Kohn-Sham Hamiltonian. From the knowledge of
$S^0_{\textbf{G},\textbf{G}'}(\textbf{q},\omega)$, the imaginary
part of $\chi^0_{\textbf{G},\textbf{G}'}(\textbf{q},\omega)$ is
easily evaluated using the relation
\begin{equation} \label{s_chi}
S^0_{\textbf{G},\textbf{G}'}(\textbf{q},\omega)
=-\frac{1}{\pi}sgn(\omega){\rm
Im}[\chi^0_{\textbf{G},\textbf{G}'}(\textbf{q},\omega)],
\end{equation}
where $sgn(\omega) = 1\ \ (-1)$ for $\omega > 0\ \ (\omega < 0)$.
The real part  of
$\chi^0_{\textbf{G},\textbf{G}'}(\textbf{q},\omega)$ is obtained
from the corresponding imaginary part using the Hilbert
transformation.

Inclusion of the $\textbf{G}\neq\textbf{G}'$ matrix elements
in Eq. (\ref{chi})
%the dielectric response couples the contributions of
%$\textbf{q}+\textbf{G}\neq\textbf{q}+\textbf{G}'$ that appears as a
%consequence of the existence of variations of the electronic density
%in real solids, see Eq. (\ref{str_fac}). This coupling
accounts for the so-called
crystalline local-field effects (LFEs)\cite{adpr62} which can be
significant if there is notable spatial variation in the electron density
in the system. If LFEs are neglected, the energy-loss function is
simply given by

\begin{multline} \label{noLFE}
\rm{Im}[\epsilon^{-1}_{\bf{G},\bf{G}}(\bf{q},\omega)]=
\dfrac{\rm{Im}\left[\epsilon_{\textbf{G},\textbf{G}}(\textbf{q},\omega)\right]}{|\epsilon_{\textbf{G},\textbf{G}}(\textbf{q},\omega)|^2}
=
\\ = \dfrac{\rm{Im}\left[\epsilon_{\textbf{G},\textbf{G}}(\textbf{q},\omega)\right]}
{
\{\rm{Re}\left[\epsilon_{\textbf{G},\textbf{G}}(\textbf{q},\omega)\right]\}^2
+
\{\rm{Im}\left[\epsilon_{\textbf{G},\textbf{G}}(\textbf{q},\omega)\right]\}^2}.
\end{multline}

In the present work, in the density functional theory (DFT) ground state calculations,
the electron-ion interaction  is represented by a norm-conserving
non-local pseudopotential,\cite{bacheleprb82} and the local density
approximation (LDA) is chosen for the exchange and correlation
potential, with the use of the Perdew-Zunger \cite{pezuprb81}
parametrization of the XC energy of Ceperley and
Alder.\cite{cealprl80} Well-converged results for the
face-centered cubic bulk Pb with the experimental lattice parameter
$a=4.95$ ${\rm \AA}$ have been obtained with a
kinetic energy cut-off of 14 Ry, including $\sim150$
plane waves in the expansion of the Bloch states.

Two different sets of calculations were carried out in evaluating
Eq. (\ref{str_fac}). First,
$S^0_{\textbf{G},\textbf{G}'}(\textbf{q},\omega)$ was calculated in
the range 0 $< \omega <$ 30 eV with a step of $\Delta\omega =$ 0.005
eV, the band indexes in Eq. (\ref{str_fac}) running up to $n =$ 25.
A Monkhorst-Pack\cite{mp} 96$\times$96$\times$96 grid of {\bf k} vectors was
used in the BZ sampling which corresponds to inclusion of $\approx$
19000 points in the irreducible part of the BZ (IBZ). The delta
function was represented by a Gaussian of width of 0.05 eV. Second,
in order to properly describe the acoustic-like modes at energies below 1 eV,
Eq. (\ref{str_fac}) was evaluated in the 0 $< \omega <$ 4 eV range,
with a step $\Delta\omega =$ 1 meV and taking into account up to 12
energy bands. In this second set of calculations a fine
432$\times$432$\times$432 grid was used, with $\approx$ 850000 {\bf
k} points in the IBZ, and the width of the Gaussian replacing the
delta function was set to 2 meV.

\begin{figure}[t]
\includegraphics[width=0.48\textwidth]{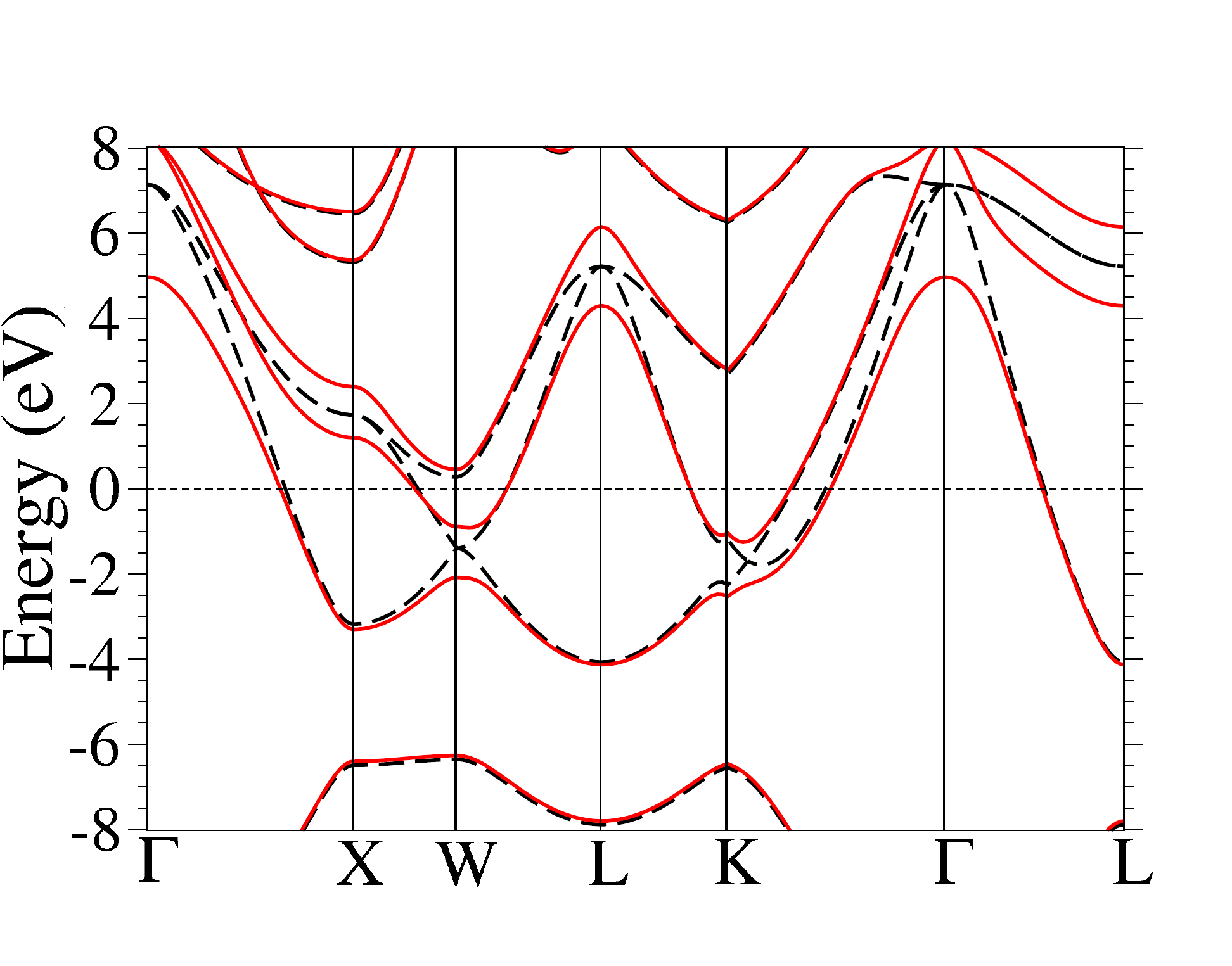}
\caption{(Color online) Band structure of bulk lead, calculated with
(solid lines) and without (dashed lines) spin-orbit coupling term in
the Hamiltonian. The horizontal dashed line represents the Fermi
level set to zero.}\label{bs}
\end{figure}

In Fig.~\ref{bs} the calculated band structure of bulk fcc lead
along some high-symmetry directions of the  first BZ is shown, with
(solid lines) and without (dashed lines) inclusion of SO interaction
in the Hamiltonian of the system. As the fcc lattice is
centrosymmetric, due to the Kramers degeneracy \cite{tinkam71} each
band is double-degenerated in both cases. The calculated band
structure is in good agreement with other calculations (see, i.e.,
Refs. \onlinecite{heboprb10,vetoprb08}) and with the experimental one
\cite{jepoprb90} when the SO term is taken into account. As can be
seen in Fig.~\ref{bs}, the inclusion of the SO interaction has
sizeable effects on the bands crossing the Fermi level (of
$p$-orbital character), mainly around the high-symmetry points of
the BZ. Thus, SO effects on the band structure of bulk Pb are
expected to show up on the low-energy dielectric properties of this
material.

The band structure calculations were performed with inclusion of the
SO term in the Hamiltonian fully self consistently. In the evaluation of
$S^0_{\textbf{G},\textbf{G}'}(\textbf{q},\omega)$, the SO coupling
enters Eq. (\ref{str_fac}) through the energy spectrum (via the
$\delta$ function) and coupling matrices (the brackets).
% in Eq. (\ref{str_fac})) were kept the same (the ones with
%the wavefunctions evaluated at the scalar-relativistic level) in
%both kind of calculations. Since the SO-split band structure of bulk
%Pb whose lattice possesses an inversion symmetry, is reproduced in
%the first order perturbation theory,\cite{chkoss91} we only expect a
%weak modification of scalar-relativistic wave function by the SO
%coupling.
We checked that the replacement of the spinor wave functions by a
scalar ones in the evaluation of $\chi^{\rm o}$ slightly modifies
the calculated dielectric properties of Pb in comparison with the
full SO results, however without important changes.

\section{GENERAL RESULTS}

In the FEG model, the plasma frequency is determined as $\omega_p =
\sqrt{4\pi n/m^*}$ with $n$ being the average electron density and
$m^*$ electron effective mass, which in terms of the density
parameter $r_s$ reads as $\omega_p = \sqrt{3/r_s^3\,m^*}$. For lead,
using the value of $r_s=2.298$ determined on base of experimental
data, one obtains $\omega_p^{\rm FEG} = 13.53$ eV. This value is in
good agreement with the one obtained in electron energy-loss
experiments (see, e.g., Ref. \onlinecite{ashtonjpf73} and references therein).
Thus, the energy transfer range $\omega\leqslant$8 eV, which is of
interest here, is well below the bulk plasmon energy in lead. Note
also that by calculating the dielectric response up to 30 eV the
present results are well converged with respect to the finite energy
range used in the numerical Hilbert transformation procedure.

\begin{figure}[t]
\includegraphics[width=0.35\textwidth,angle=270]{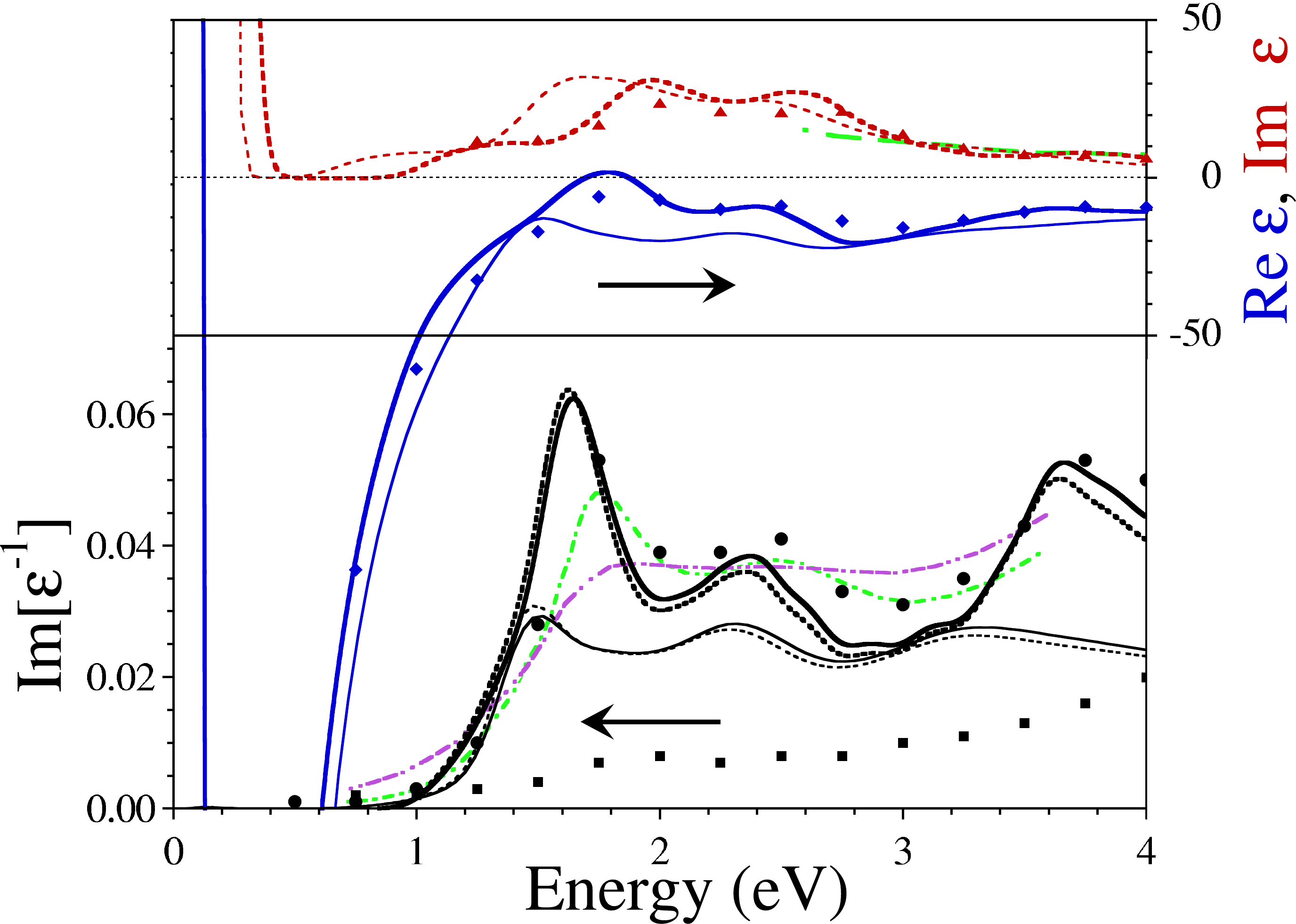}
\caption{(Color online) (Upper panel) Imaginary (dashed lines) and
real (solid lines ) parts of the dielectric function obtained in the RPA
calculations at Q=0.014 a.u. along the $\Gamma$-X direction. In the
lower panel the corresponding energy-loss function evaluated with (solid
lines) and without (dotted lines) inclusion of the LFEs is shown.
Thick (thin) lines present results obtained with (without) inclusion
of spin-orbit splitting.  Experimental data for the loss function from
Ref. \onlinecite{mathewps71} measured at 140 K and room temperature
are shown by thin dashed-dotted and dashed-dashed-dotted-dotted
lines, respectively. Thick long-dashed line in the upper panel shows
the measured imaginary part of the dielectric function.\cite{leprprb73}
Filled triangles, diamonds, and circles present corresponding data
obtained in first-principles calculations of Glantschnig and
Ambrosch-Draxl.\cite{wegljpcrd09,glamnjp10} Filled squares show loss
function obtained in the REELS experiment.\cite{wegljpcrd09}}
\label{optics}
\end{figure}

\begin{figure}[t]
\includegraphics[width=.48\textwidth]{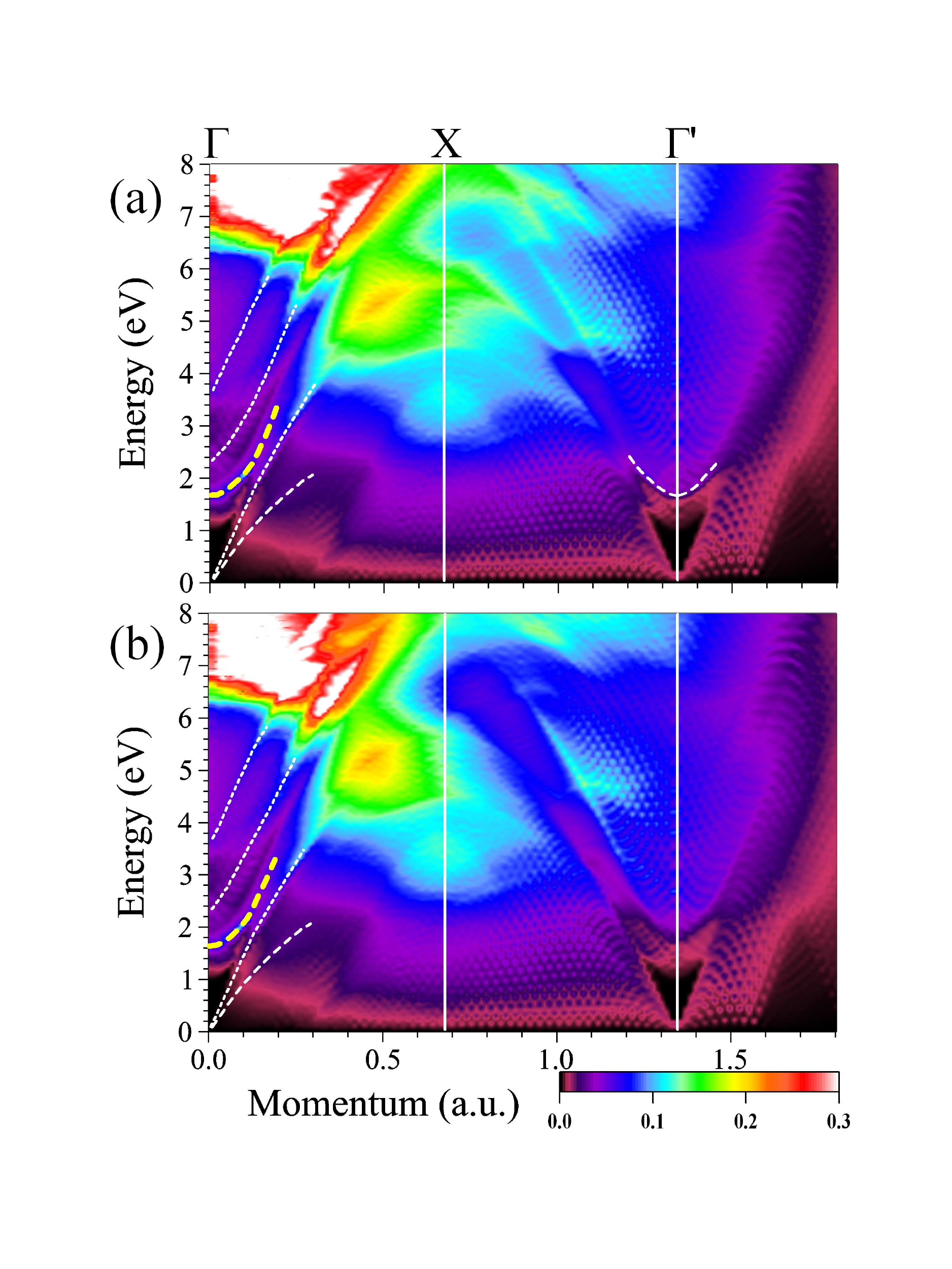}
\caption{(Color online) Calculated energy-loss function of Pb versus
energy $\omega$ and momentum transfer $Q$ along the $\Gamma$-X
symmetry direction. Results are obtained with inclusion of the SO
coupling and the RPA kernel and (a) with and (b) without inclusion
of the LFEs. Thick yellow dashed lines highlight dispersion of the
plasmon modes. Thin white dashed lines show peaks corresponding to
strongly damped modes, while the thin dotted ones highlight the modes which
strongly disperse upwards up to energies $\omega \gtrsim$ 4 eV. Vertical white lines mark positions of the X
point and the $\Gamma$ point in the subsequent BZ. } \label{Dir_100}
\end{figure}

\begin{figure}[t]
\includegraphics[width=.48\textwidth]{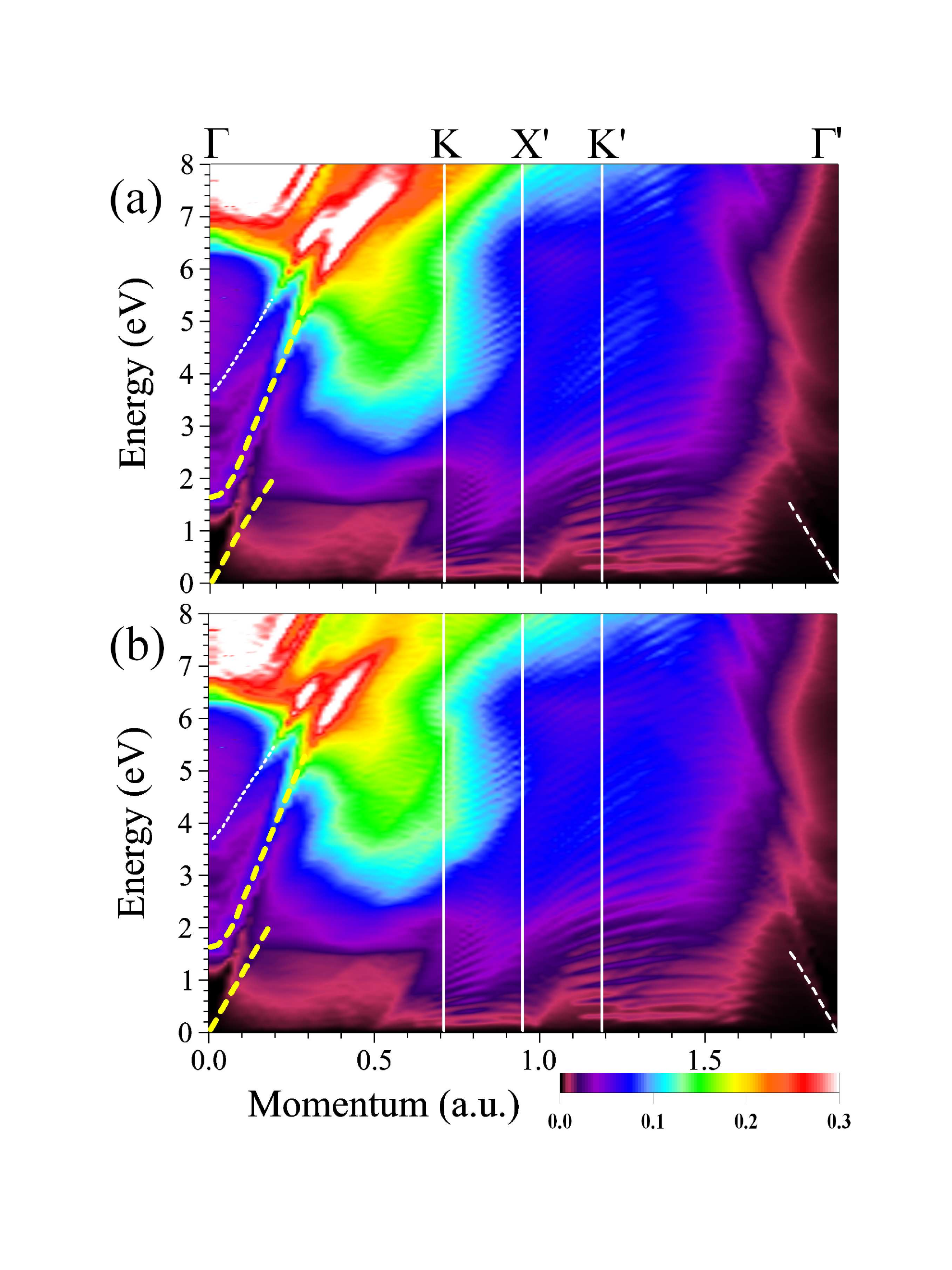}
\caption{(Color online) Calculated energy-loss function of Pb versus
energy $\omega$ and momentum transfer $Q$ along the $\Gamma$-K
symmetry direction. Results are obtained with inclusion of the SO
coupling and the RPA kernel and (a) with and (b) without inclusion
of the LFEs. Thick yellow dashed lines highlight dispersion of the
plasmon modes. Thin white dashed lines show peaks corresponding to
strongly damped modes, while the thin dotted one highlights an
interband mode dispersing upwards until it enters the manifold of s-p transitions. Vertical white lines mark positions of the K
point and the X, K, and $\Gamma$ points in the subsequent BZs.}
\label{Dir_110}
\end{figure}

\begin{figure}[t]
\includegraphics[width=.48\textwidth]{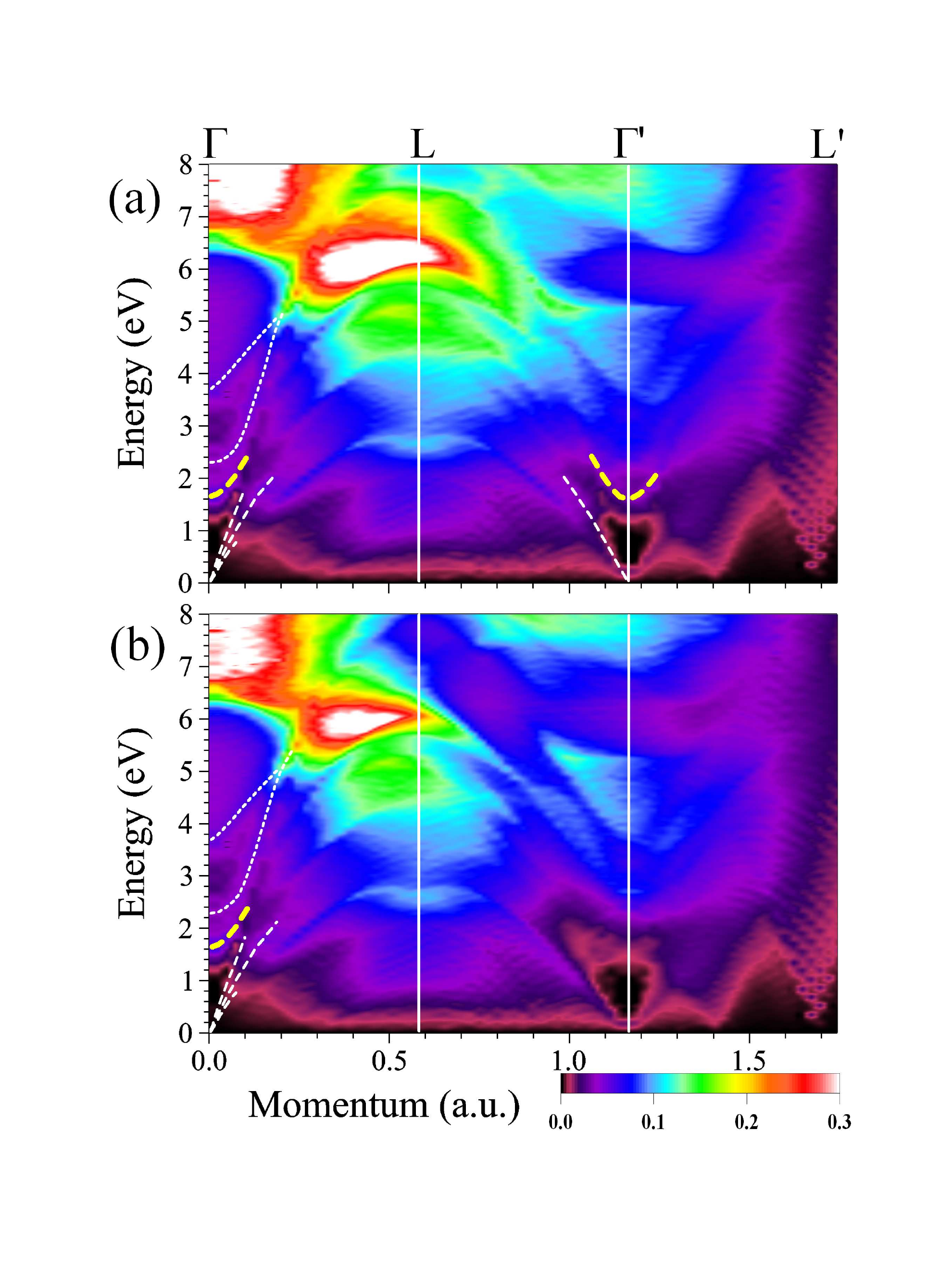}
\caption{(Color online) Calculated energy-loss function of Pb versus
energy $\omega$ and momentum transfer $Q$ along the $\Gamma$-L
symmetry direction. Results are obtained with inclusion of the SO
coupling and the RPA kernel and (a) with and (b) without inclusion
of the LFEs. Thick yellow dashed lines highlight dispersion of the
plasmon modes. Thin white dashed lines show peaks corresponding to
strongly damped modes, while the thin dotted ones highlight
two interband modes dispersing upwards until they merge each other. Vertical white lines mark positions of the L
point and the $\Gamma$ and L points in the subsequent BZs.}
\label{Dir_111}
\end{figure}

\begin{figure}[t]
\includegraphics[width=0.48\textwidth,angle=0]{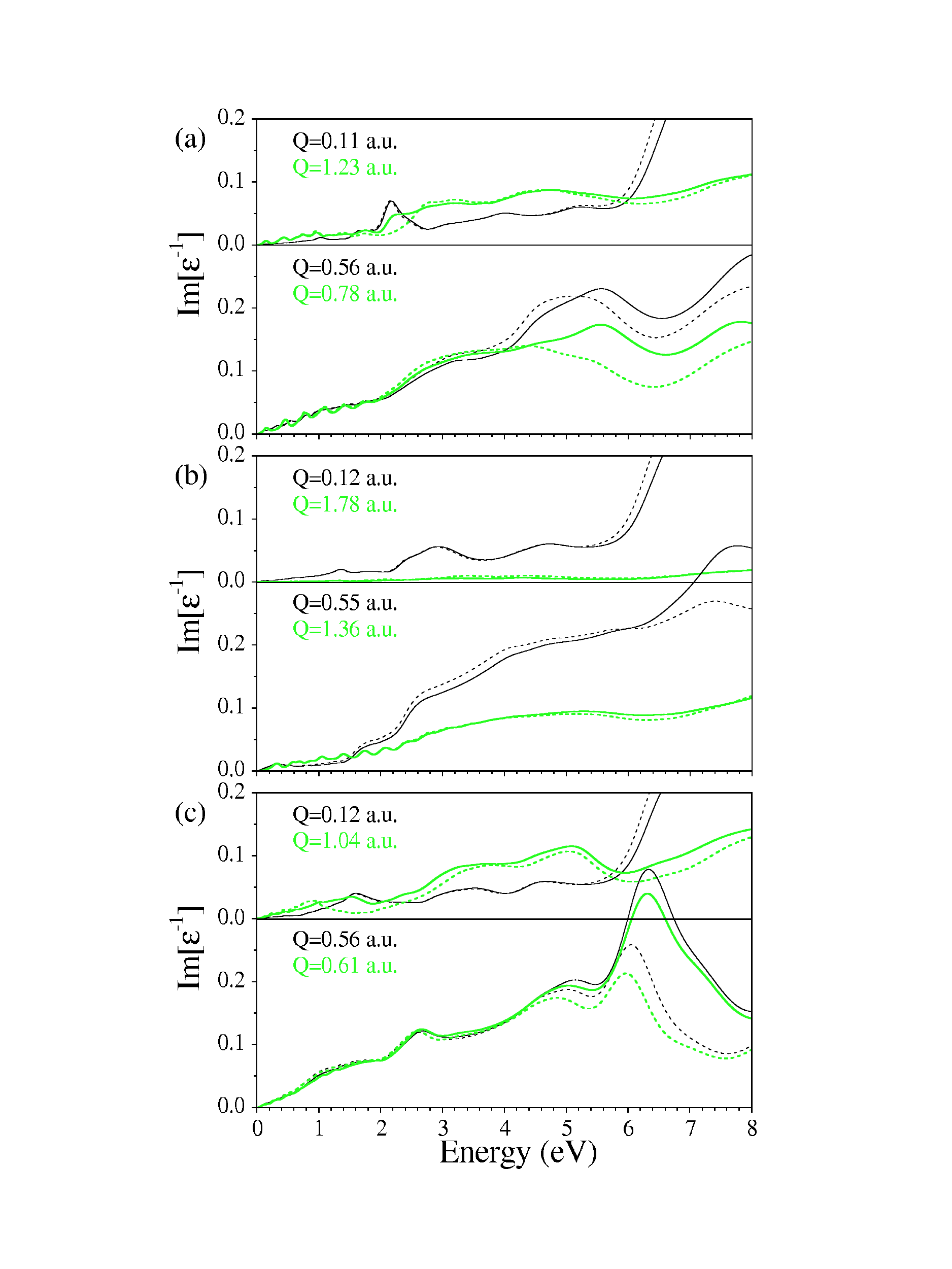}
\caption{(Color online) Calculated energy-loss function at some
fixed momentum transfers Q's as a function of energy. Calculations
are performed with inclusion of the SO interaction at the RPA level.
Solid (dashed) lines are obtained with (without) inclusion of LFEs.
Thick (thin) lines stand for the greater (smaller) Q reported in each panel.
The data at \textbf{Q}'s along $\Gamma$-X, $\Gamma$-K, and
$\Gamma$-L symmetry directions are presented in (a), (b), and (c),
respectively.} \label{LFE}
\end{figure}

Comparison of the calculated energy-loss function with optical
experimental data from Ref. \onlinecite{mathewps71} obtained at 140
K and room temperature is performed in Fig.~\ref{optics}. The
calculated curves correspond to the smallest momentum transfer
$Q$=0.014 a.u. along the $\Gamma$-X direction, allowing comparison
with optical measurements. As can be seen in the figure, the loss
function calculated without the spin-orbit interaction significantly
deviates from both experimental curves. In particular, the first
peak in the calculated curve is located at 1.5 eV, i.e. at a notably
lower energy in comparison with experiment. Other broad peaks
centered at energies of 2.35 eV and 3.4 eV, are also located at
lower energies. Only inclusion of the SO coupling in the
band structure leads to a fairly good agreement in the energy positions of
all three features in this energy range with the experimental ones,
especially with the measurements carried out at 140 K. Also our data
with the SO interaction included are in good agreement with
{\it ab initio} calculations of Ref. \onlinecite{glamnjp10}.
Some quantitative differences observed between two {\it ab initio} results are at the level of uncertainty in such kind of calculations.

In Figs.~\ref{Dir_100}-\ref{Dir_111} the energy-loss function,
Im$\left[\varepsilon^{-1}(\textbf{Q},\omega)\right]$,
%\equiv{\rm
%Im}\left[\epsilon^{-1}_{{\bf G},{\bf G}}(\textbf{q},\omega)\right]$,
calculated using the RPA approximation for the exchange-correlation
kernel and including the SO interaction, is presented as a
function of the energy transfer $\omega$ and value of the momentum
transfer \textbf{Q} along three high-symmetry directions, both
including and neglecting the LFEs. In these figures, at small $Q$'s
one can observe several peaks in the energy range below the
prominent broad peak structure presented at energies above $\sim
6.5$ eV which coincides with the energy threshold for interband
transitions between the occupied $s$ states (those located at
energies below -6 eV in Fig.~\ref{bs}) and the $p$ states above the
Fermi level.

The strong increase of intensity of the interband peak at $\sim$1.65 eV in the calculated loss function with the SO coupling included can be explained by the fact that, as seen in the upper panel of Fig.~\ref{optics}, the real part of the dielectric function reaches zero at a close energy due to a sharp increase in the corresponding imaginary part of the dielectric function. It seems in our calculations this effect is more pronounced than in the calculation of Glantsching and Ambrosch-Draxl.\cite{glamnjp10} Nevertheless, the evaluated loss functions in both calculations are rather similar in this energy range. Only little differences can be observed at larger energies, where the loss function  calculated in Ref. \onlinecite{glamnjp10} is slightly larger than that evaluated in the present work. This is explained by the larger imaginary part of the dielectric function in our calculation in the 1.7-2.7 eV energy region and the lower real part at energies above 2 eV. From this analysis we can conclude that the experimental optical data reflect clear SO effects. Note that in the experimental data of Ref. \onlinecite{mathewps71} the two first lowest energy peaks are barely
visible in the  room temperature loss spectra and significantly more
pronounced in the measurements performed at T=140 K. This is in
agreement with the fact that Pb presents a strong electron-phonon
coupling, which modifies the one-electron energy levels with
increasing temperature. Hence, one can expect that in measurements
performed at even lower temperatures, the first interband peak might
increase its intensity and downshift in energy, in such a way
improving agreement with the calculations.

On the other hand, comparison of the calculated results with the
recently measured data obtained by reflection electron energy-loss
spectroscopy (REELS)\cite{wegljpcrd09} reveals only a broad weak
peak around 2 eV and presents significant underestimation in
intensity of the whole spectrum. These might be a consequence that
this REELS experiment setup might have insufficient resolution in
this low-energy range, being apparently more suitable at higher
energies.\cite{wegljpcrd09}

In Figs.~\ref{Dir_100}-\ref{Dir_111} one can see how the peaks in
the loss function strongly disperse upward upon increase of momentum
transfer values in all three directions. Thus the dominating 1.65 eV
peak increases its energy up to $\sim3.5$ eV ($\sim8$ eV) at $Q=0.2$ $(0.6)$
a.u. along the $\Gamma$-X ($\Gamma$-K) direction. On the other hand, the
1.65 eV possesses along the $\Gamma$-L direction much less dispersion and quickly disappears at
$\omega=2.3$ eV.
The other weaker 2.35 eV and
3.7 eV peaks disperse upward in $\Gamma$-X and $\Gamma$-K up to energies
above 6 eV where they enter the manifold corresponding to $s-p$
interband transitions and can be resolved as separate features up to
$\omega=8$ eV. On the contrary, along $\Gamma$-L these upper-energy peaks disperse up to
energy of $\sim5.2$ eV where they merge each other and a much stronger peak continues
dispersion up to energies about 6.3 eV. Starting from $Q=0.6$ a.u. the
dispersion of this peak turns from positive to negative and it can
be clearly resolved up to $Q\sim1$ a.u.

Below we present a systematic analysis of the effect of the main
physical ingredients like LFEs, SO coupling, and XC effects on the
low-energy electronic collective excitations in bulk Pb.

\subsection{Local-field effects}

From comparison of the upper and lower panels in Figs.~\ref{Dir_100}-\ref{Dir_111} one can deduce that the LFEs affect the
calculated dielectric properties rather weakly. From Figs.~\ref{optics} and \ref{LFE} it is seen that this effect is barely
visible at small momentum transfers. Along the $\Gamma$-X,
$\Gamma$-K and $\Gamma$-L directions the main result of the LFEs in
the formation of the excitation spectra in Pb is some distortion of
the intensity of the aforementioned peaks at the finite momentum
transfers. At energies above $\omega\sim6$ eV the LFEs produce
significant increases in intensity of the dominant peaks.

\begin{figure}[b]
\includegraphics[width=0.48\textwidth]{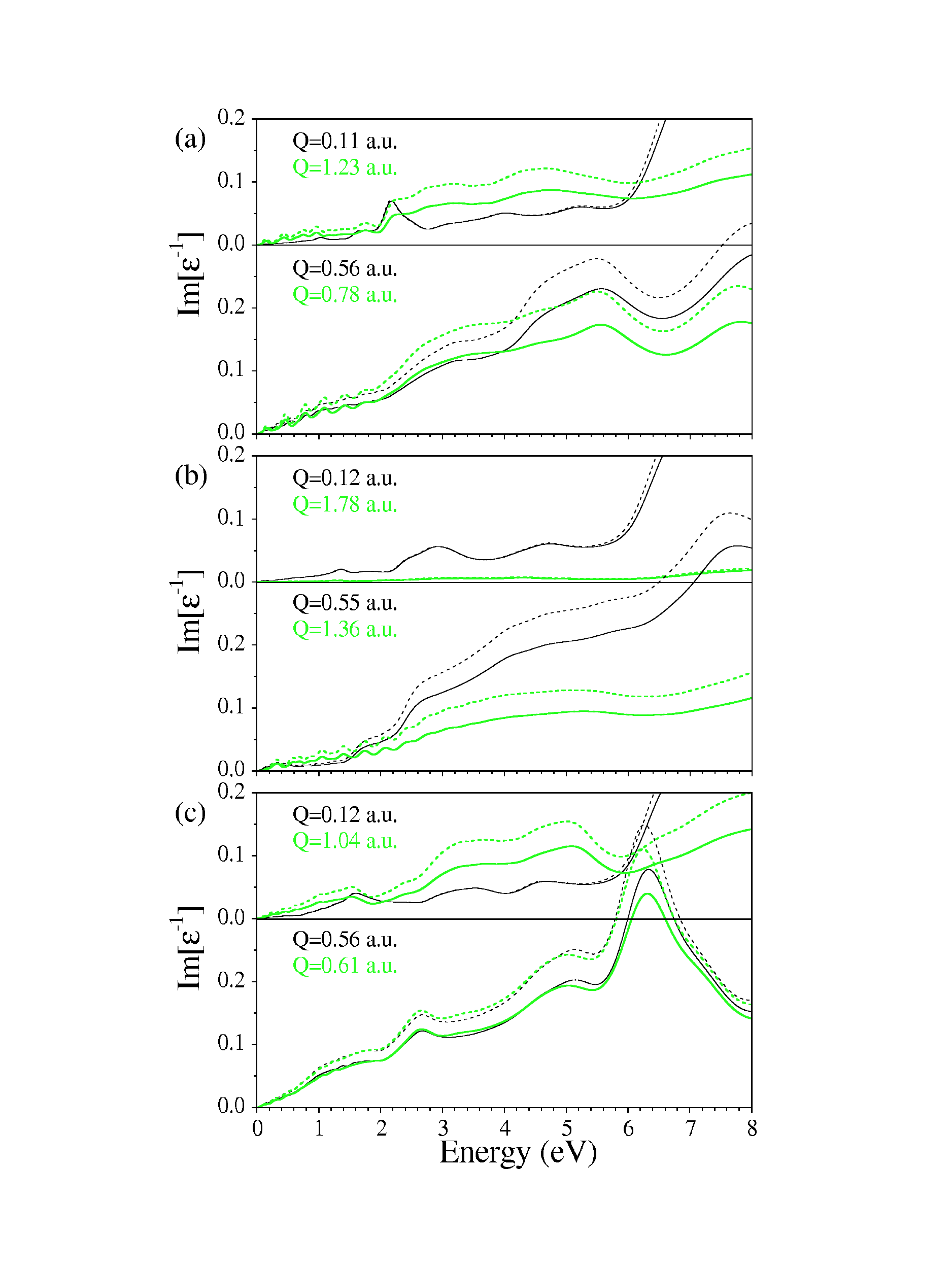}
\caption{(Color online) Solid lines represent the same as in Fig.~\ref{LFE}, while the dashed lines show the energy-loss function
evaluated at corresponding \textbf{Q}'s with inclusion of XC effects at the TDLDA level.
Thick (thin) lines stand for the greater (smaller) Q reported in each panel.
}\label{TDLDA}
\end{figure}

Additionally the LFEs produce an upward shift in energy of all the
features. However the effect is not very pronounced and in general
do not exceed several tenths of eV. Thus in Fig.~\ref{LFE} one can
see that the major upward shift about 0.3-0.4 eV occurs in the case
of the $\sim6$ eV feature at intermediate momentum transfers along
the $\Gamma$-L direction.

Another consequence of the inclusion of the LFEs is transmission of the
1.65 eV peak at small $Q$'s to momentum transfers close to
$Q=2\pi/a=1.35$ a.u. in the $\Gamma$-X direction. At momentum
transfers close to this $Q$ one can see how intensity of the loss
function at $\omega\approx1.65$ eV notably increases when the LFEs
are included. At $Q=1.23$ a.u. the corresponding increase in the
loss function at $\omega\approx2.25$ eV due to LFEs can be seen in
Fig.~\ref{LFE}(a). Also some increase in the loss function caused by
the LFEs due to the $\omega=1.65$ eV mode can be detected in the vicinity of $Q=1.16$ a.u. along the $\Gamma$-L direction. The
example of this enhancement can be seen in Fig.~\ref{LFE}(c), where
a broad feature in the loss function appears at energies around 1.6
eV at $Q=1.04$ a.u. when the LFEs are taken into account. However,
in general, the impact of the LFEs on the loss spectra in Pb, being
noticeable at certain energies, is not so strong as in other systems like MgB$_2$\cite{silkinprb09a} and compressed
lithium.\cite{erroprb10} This signals about a less inhomogeneity in
the valence charge density in Pb in comparison with those systems.

\subsection{XC kernel}

In Fig.~\ref{TDLDA}, the calculated energy-loss function for several
values of \textbf{Q} belonging to the three different high-symmetry
directions is plotted, where comparison between results obtained
with the RPA and the TDLDA kernels is made. As can be seen in
the figure, the main effect of the TDLDA with respect to the RPA is
the increase of the intensity of the calculated
$\rm{Im}\left[\varepsilon^{-1}(\textbf{Q},\omega)\right]$, but without
qualitative changes in its shape. The most significant change is
seen in Fig.~\ref{TDLDA}(c) where the dominant interband peak in the
$\Gamma$-L direction is shifted downward in energy by $\sim$ 0.1 eV
upon inclusion of the XC effects at the TDLDA level.

\subsection{SO-interaction-induced effects}

\begin{figure}[t]
\includegraphics[width=0.48\textwidth]{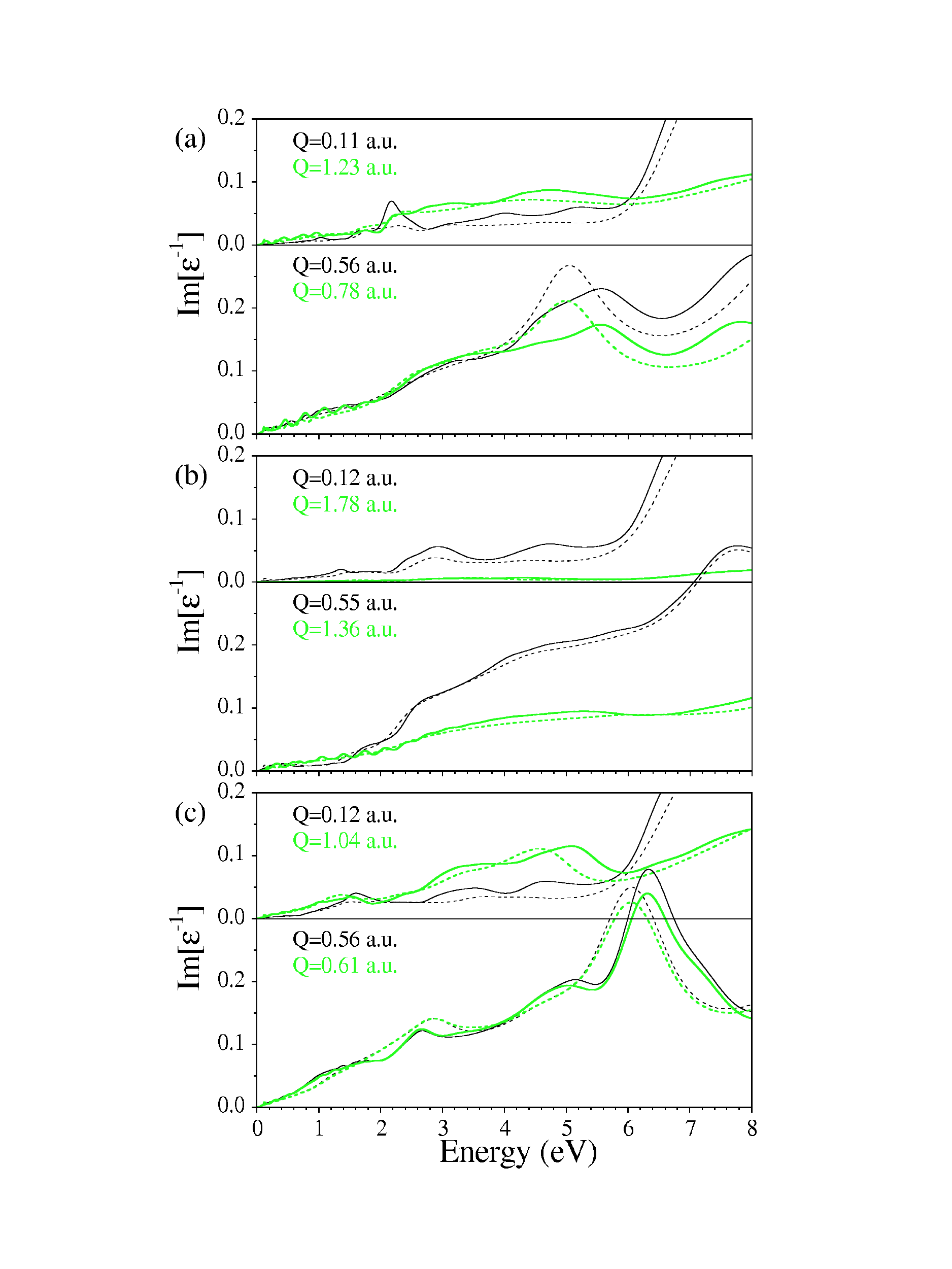}
\caption{(Color online) Solid lines represent the same as in Fig.~\ref{LFE}, while the dashed lines show the energy-loss function
evaluated at corresponding \textbf{Q}'s at the RPA level without
inclusion of the SO interaction. Thick (thin) lines stand for the greater (smaller) Q reported in each panel.
}\label{SO}
\end{figure}

Inclusion of the SO coupling also affects the dielectric
response of bulk Pb in an anisotropic way.  As shown in Fig.~\ref{SO}, in all three high-symmetry directions the SO interaction
increases the intensity of the broad feature located at energies above
6 eV. As was previously discussed inclusion of the SO in
the calculation of the energy-loss function has sizeable effects at
small momentum transfers at energies of 1.65 eV, 2.35 eV, and 3.7
eV. The impact of the SO coupling on the excitation spectra can also
be observed in Fig.~\ref{SO}. Thus at $Q=0.11$ a.u. along
$\Gamma$-X the appearance of a clear peak at 2.15 eV and two broad
peaks at $\omega=4.0$ eV and $\omega=5.2$ eV caused by the SO
interaction can be appreciated. At larger $Q$'s the effect is
smaller and consist mainly in the upward shift of the existing
peaks. The same trend is observed in the $\Gamma$-K and $\Gamma$-L
directions as well, although with less impact at intermediate and
large momentum transfers. Additionally, in Fig.~\ref{SO} one can
detect that at small $Q$'s in the low-energy region the inclusion of
the SO interaction leads to the appearance of a pronounced peak. Thus
at $Q=0.11$ a.u. along the $\Gamma$-X direction it is located at
$\omega=1$ eV, whereas at $Q=0.12$ a. u. the corresponding peak
presents at $\omega=1.15$ eV and $\omega=1.6$ eV along the
$\Gamma$-K and $\Gamma$-L directions, respectively. The dispersion
of these peaks is highlighted by dashed lines in Figs.~\ref{Dir_100}-\ref{Dir_111}
and it is discussed in the next section.

\section{ACOUSTIC-LIKE EXCITATIONS}

\begin{figure}
\centering
\includegraphics[width=0.48\textwidth]{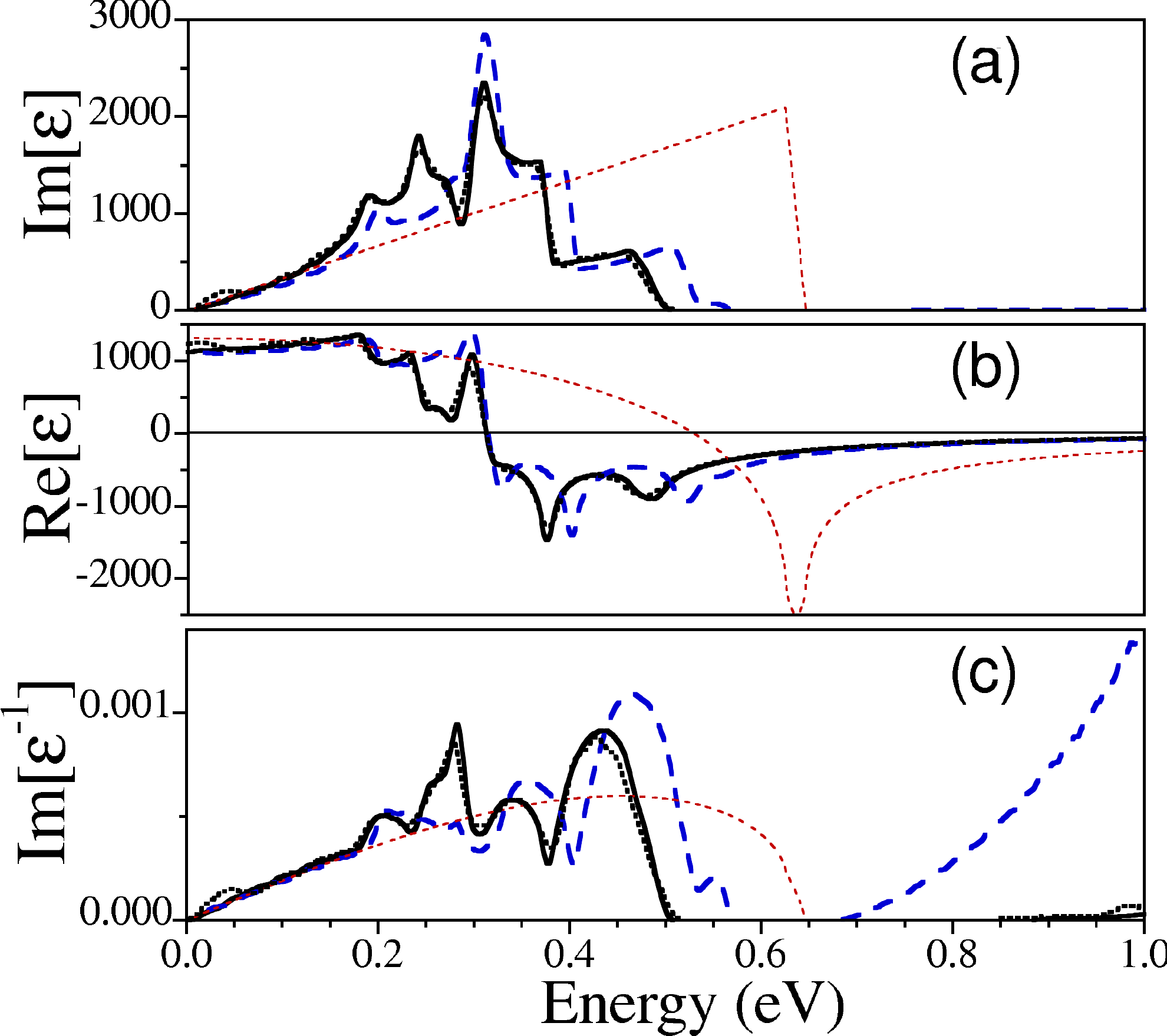}
\caption{(Color online) Calculated (a) imaginary and (b) real parts of dielectric function and (c) energy-loss function evaluated at $q\simeq0.028$ a.u. along the $\Gamma$-X direction.
Thick dashed, dotted, and solid lines correspond to the results obtained at the scalar-relativistic level, with inclusion of the SO coupling and scalar representation for wave functions, with inclusion of the SO coupling and spinor representation for wave functions, respectively. Thin dotted line shows corresponding quantities obtained with the Lindhard dielectric function for $r_s^{\rm Pb}=2.298$. {\it Ab initio} results are obtained with inclusion of the LFE and TDLDA kernel.
}\label{fig:APX}
\end{figure}

\begin{figure}
\centering
\includegraphics[width=0.48\textwidth]{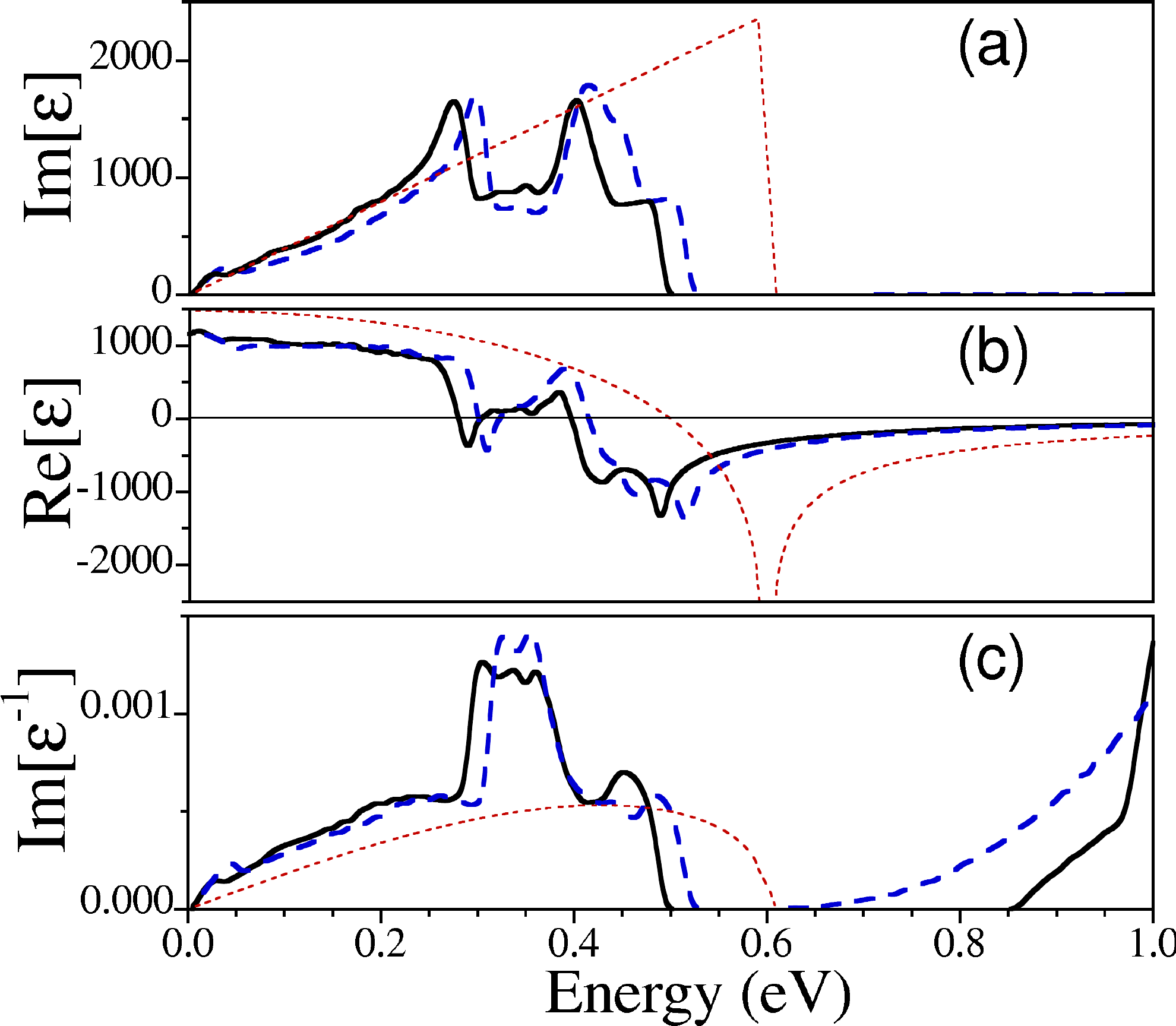}
\caption{(Color online) The same as in Fig.~\ref{fig:APX} evaluated at $q=0.026$ a.u. along the high-symmetry $\Gamma$-K direction.}\label{fig:APK}
\end{figure}

\begin{figure}
\centering
\includegraphics[width=0.48\textwidth]{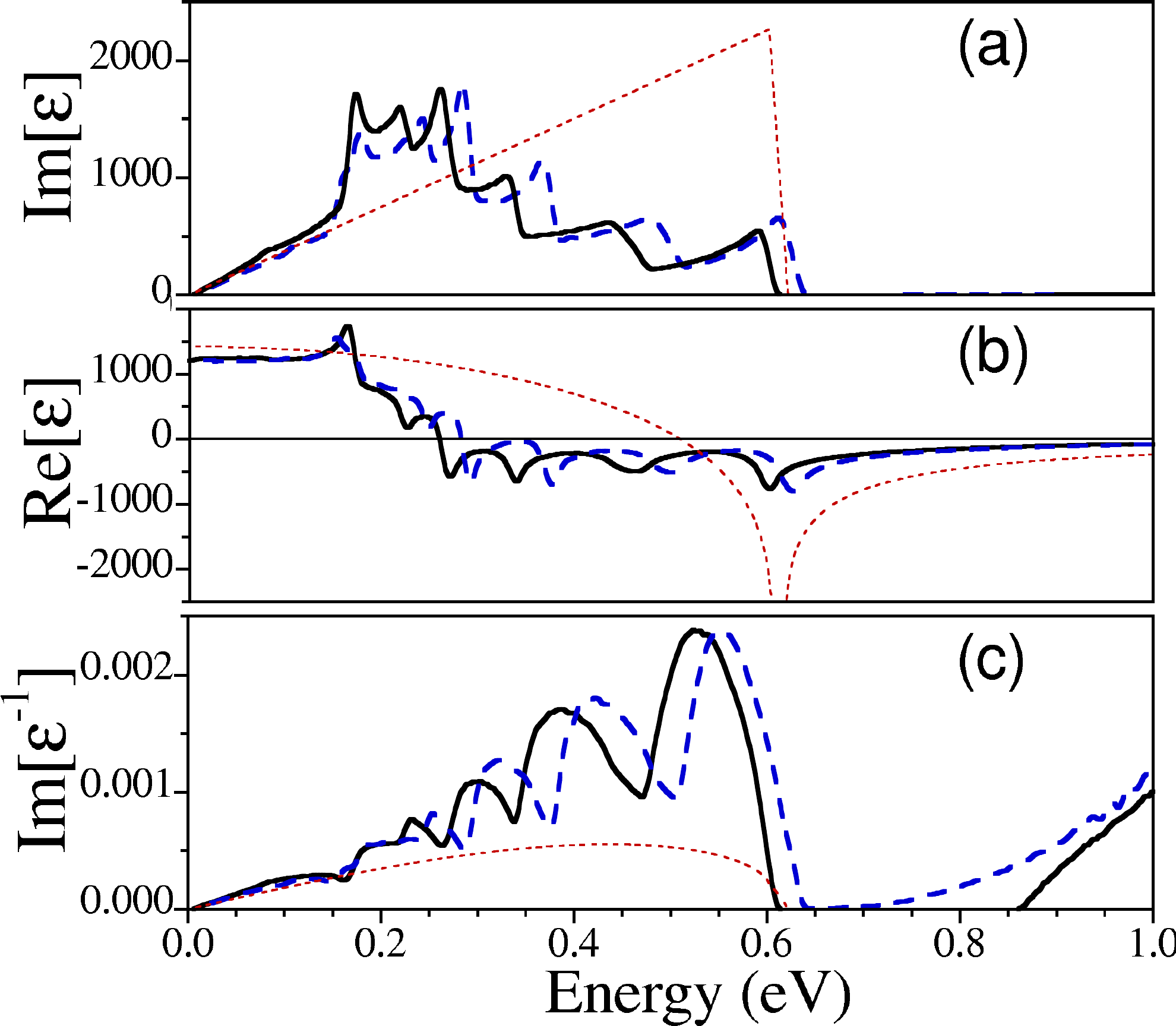}
\caption{(Color online) The same as in Fig.~\ref{fig:APX} evaluated for $q=0.027$ a.u. along the high-symmetry $\Gamma$-L direction.}\label{fig:APL}
\end{figure}

In the \textquotedblleft low momentum--low energy\textquotedblright\, region of Figs.~\ref{Dir_100}-\ref{Dir_111} the calculated loss function presents several peaks dispersing almost linearly with momentum, whose energy is vanishing upon vanishing of the momentum transfer. Upon momentum transfer increase these peaks can be traced up to an energy of about 2 eV in $\Gamma$-K and $\Gamma$-L, and up to even higher energies in $\Gamma$-X. The number of such peaks depend on the direction, being two in the case of $\Gamma$-X, one in $\Gamma$-K, and three along $\Gamma$-L. To study in more detail the origin of these modes characterized by an acoustic-like dispersion, in Figs.~\ref{fig:APX}-\ref{fig:APL} we report the calculated dielectric and energy-loss functions at almost the same small momentum transfers in all three high-symmetry directions.
Here we show the results obtained at the scalar-relativistic level and with inclusion of the SO coupling and spinor representation for wave functions. For comparison, in these figures the featureless curves derived from the Lindhard dielectric function\cite{grosso}  for the same q's are presented as well. The {\it ab initio} curves were obtained with inclusion of the LFE and the TDLDA kernel, even though these two physical ingredients were found to affect negligibly the results in this low-energy range in all cases.

In Fig.~\ref{fig:APX} one can observe two clear peaks in the loss function, whose shape and intensity depends on whether the SO interaction is included or excluded from the calculation whereas spinor representation for the wave functions has minor impact on the results as was suggested in Ref. \onlinecite{zubizarreta}. Note that we observe a similar little impact on the dielectric properties of Pb of the spinor representation for the wave functions in other symmetry directions as well. From the analysis of the real part of the dielectric function we conclude that neither of these peaks can be considered as a true plasmon mode as the real part of $\varepsilon$ does not cross zero at the corresponding energies. The only zero-crossing in Re$[\varepsilon]$ occurs at 0.32 eV close to the energy where Im$[\varepsilon]$ has a maximum. As a result, at this energy the loss function presents a local minimum. Therefore this zero-crossing must be considered as a conventional Landau-overdamped mode which can not be realized.\cite{pino66} At the same time, inspection of Fig.~\ref{fig:APX} shows that the peaks presented in the energy-loss function are located at energies where the Im$[\varepsilon]$ possesses local minima. Hence, despite rather large values of the Re$[\varepsilon]$ at corresponding energies these peaks can be considered as corresponding to heavily damped acoustic plasmons. One can compare these results with those derived in the FEG model, where a zero-crossing of Re$[\varepsilon]$ does not produce any peak in the loss function due to the peak in Im$[\varepsilon]$ at the close energy.

The presence of peaks in the loss function is explained by the presence of a number of peaks in Im$[\varepsilon]$. These peaks in Im$[\varepsilon]$ are due to intraband excitations within the energy bands crossing the Fermi level. Although all these bands are of the same $p$-like character, their dispersion with different Fermi-velocity components in this symmetry direction is reflected in the presence of several peaks in Im$[\varepsilon]$.

Concerning the $\Gamma$-K direction, the loss function presented in Fig.~\ref{fig:APK} shows a broad main peak which is centered at $\sim$0.33 eV when the SO coupling is included in the calculations. Moreover, one can see that Re$[\varepsilon]$ is rather small at that energy and even crosses zero at 0.30 eV with positive slope, when SO is included in the calculations. This signals that this peak corresponds to a true plasmon mode, although severely damped due to decay into electron-hole pairs. The presence of this peak in Im$[\varepsilon^{-1}]$ can be explained by the presence of two clear main peaks in Im$[\varepsilon]$ at 0.26 eV and 0.40 eV, again when the SO interaction is included in the calculations. This makes the real part of the dielectric function cross zero three times, the second one with positive slope leading to the appearance of the peak in the loss function. Again, as it is in the $\Gamma$-X direction, in the scalar-relativistic case all the peaks in Im$[\varepsilon
 ]$ and Im$[\varepsilon^{-1}]$ are located at higher energies.

At momentum transfers along the $\Gamma$-L direction the number of peaks in the loss function is maximal. As an example, in Fig.~\ref{fig:APL} one can detect up to three clear peaks in Im$[\varepsilon^{-1}]$ at energies of 0.30 eV, 0.38 eV, and 0.53 eV. Their dispersion is shown in Fig.~\ref{Dir_111}. The presence of such large number of peaks in Im$[\varepsilon^{-1}]$ can be again explained by a large number of peaks in Im$[\varepsilon]$ seen in Fig.~\ref{fig:APL} (up to five). However, neither of these peaks leads to an additional zero-crossing in Re$[\varepsilon]$. For this reason, all these peaks can be considered as being heavily damped plasmonic modes. Similar to what occurs in other symmetry directions the effect of inclusion of SO coupling is limited to the downward shift of these modes without qualitative changes.

In the calculation at the scalar-relativistic level we obtain that the upper border for the intraband electron-hole transitions is located at higher energies than when the SO interaction is included. This is explained by modifications in the energy bands around the Fermi surface. The main effect of the inclusion of the SO is the flattening of the band dispersion accompanying the opening energy gaps seen in Fig.~\ref{bs}. Consequently, this causes modifications of intraband excitations reflected in the integrated form through Eq.~(\ref{str_fac}) in Im$[\varepsilon]$. Regarding the shape of the acoustic-like dispersing modes, the SO coupling only slightly affects $\varepsilon$ at \textbf{q}'s in $\Gamma$-X, where it gives rise to a new peak (which is located at $\omega\simeq0.27$ eV in Fig.~\ref{fig:APX}).

\subsection{Group velocities: comparison with $v_F$} \label{sub:4.vg}

Concerning the group velocities $v_g$ of the acoustic modes, the values are dependent on the momentum transfer direction as readily seen from the slopes of the corresponding lines in Figs.~\ref{Dir_100}-\ref{Dir_111}, thus showing anisotropy as a result of band structure effects. More precisely, the group velocities $v_g$ present values of 0.33 and 0.41 a.u. in the $\Gamma$-X direction, 0.41 a.u. in $\Gamma$-K, and 0.40, 0.51 and 0.71 a.u. in $\Gamma$-L. The velocity atomic unit is 2.1877$\times10^6$ m/s. All the reported velocities are lower than the Fermi velocity derived from the FEG model \cite{ashcroft} of $v_F^{FEG}=0.84$ a.u. On the other hand, all the estimated $v_g$ are higher than the experimental value of $v_F^{exp}=0.23$ a.u.\cite{Bardeen1961} obtained in skin depth measurements. Note the estimated group velocities of the acoustic modes can not be simply assigned to the Fermi velocities of the bands crossing the Fermi surface on a fixed reciprocal space point in the calculated band structure. As an example, comparison of the above reported values of $v_g$ with the maximal Fermi velocities of the bands in the high-symmetry directions (see Fig.~\ref{bs}) of 0.60 a.u. ($\Gamma$-X), 0.47 and 0.54 a.u. ($\Gamma$-K) and 0.84 a.u. ($\Gamma$-L) shows clear deviations between the calculated $v_F$ and $v_g$ values. In all cases $v_F > v_g$. This is as expected, since collective electronic excitations can not be built faster than the velocity of the individual electrons. Thus, in each \textbf{q} direction, the maximum $v_F$ can be seen as the upper bound for the group velocities of the acoustic modes.

\subsection{Possibility of detection in EELS experiments} \label{sub:4.APEELS}

\begin{figure}
\centering
\includegraphics[width=0.48\textwidth]{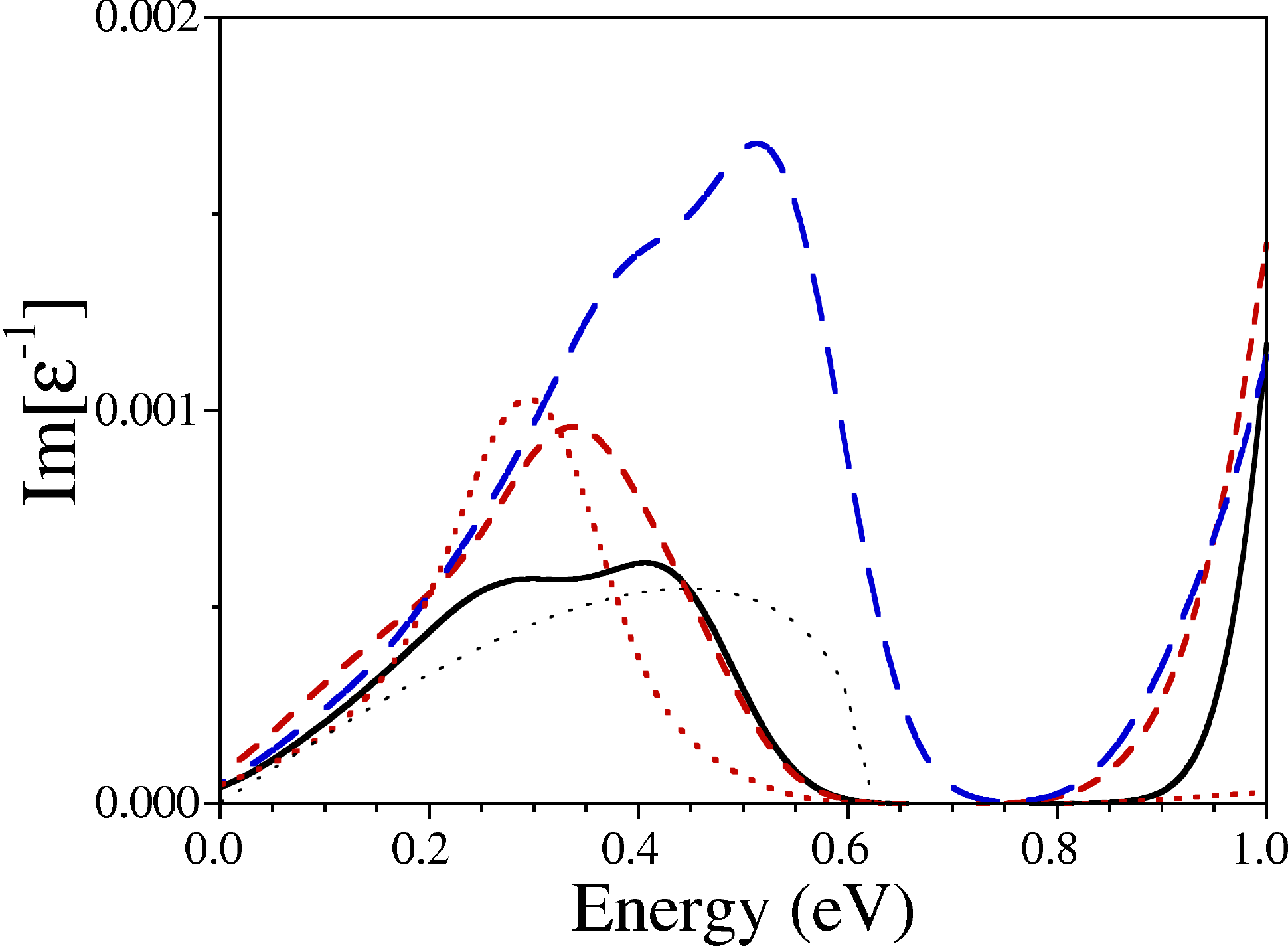}
\caption{(Color online) Comparison of the energy-loss function for the acoustic dispersing peaks in all three high-symmetry directions at the same momentum transfers as in Figs. \ref{fig:APX}-\ref{fig:APL}. Thick solid, dashed, and long-dashed lines show results at $q$'s along the $\Gamma$-X, $\Gamma$-K, and $\Gamma$-L symmetry directions, respectively. Thick dotted line is the loss function evaluated at momentum transfer corresponding to vector $q$ in the second BZ along $\Gamma$-K. Loss function derived from the Lindhard dielectric function at $q=0.027$ a.u. is presented by the thin dotted line. First-principles results are broadened with a broadening parameter of 75 meV (see text).}\label{fig:broad}
\end{figure}
In order to evaluate the possibility of experimental detection of the aforementioned acoustic modes, in Fig.~\ref{fig:broad} a comparison of the energy-loss function for the acoustic dispersing peaks at the same momentum transfers reported in Figs. \ref{fig:APX}-\ref{fig:APL} in all three high-symmetry directions, together with the loss function derived from the Lindhard dielectric function at  $q=0.027$ a.u. Also the acoustic-like plasmon mode is shown for \textbf{q} in the second BZ in the $\Gamma$-K direction by a dashed red line. A Gaussian broadening has been applied to the first-principles results. The broadening parameter was fixed as 75 meV corresponding to the experimentally measured linewidth of quantum-well states in Pb(111) thin films \cite{hobrprb09} at T = 5 K. The intensity of the acoustic mode peak is maximal in the $\Gamma$-L direction. However, from Fig.~\ref{fig:broad} we conclude that the most suitable acoustic plasmon for experimental detection corresponds to the peak dispersing in the second BZ in the $\Gamma$-K direction. It presents the smaller linewidth once the broadening is applied, and more importantly it is located in a $(q,\omega)$ range in which the energy-loss function presents vanishing values except for the acoustic mode itself, getting isolated in this way (see Fig.~\ref{fig:broad}). Note also that the cross section probed in EELS experiments is proportional to $q^{2}$,\cite{Kuzmany1998} making weak features in the low-momentum transfer range more difficult to be resolved.

However, an additional possibility of detecting an acoustic plasmon in bulk Pb can be suggested based on the general results obtained for the momentum transfer along the $\Gamma$-X direction. As seen from Fig.~\ref{Dir_100}, the fastest dispersing acoustic mode reaches an energy of $\sim$5 eV at $q\simeq0.45$ a.u., with a gradual increase of its intensity. Thus, once the peak at $\sim$5 eV and $q\simeq0.45$ a.u. is detected, one could trace its dispersion back towards vanishing energy and momentum transfer values.\\

\section{CONCLUSIONS} \label{sec:5.summary}

We have presented first-principles calculations of the low-energy ($\omega\leqslant$8 eV) electronic  collective excitations in bulk Pb and studied in detail the effect of the main physical ingredients involved, as well as the existence and character of acoustic-like modes. Good agreement with available optical experimental data \cite{mathewps71} is interpreted as an evidence of remarkable SO effects, also in agreement with other theoretical works.\cite{wegljpcrd09,glamnjp10}

In general, strong anisotropic effects are found, resulting in a distinct topology of ${\rm Im}[\varepsilon^{-1}_{\textbf{G},\textbf{G}}(\textbf{q},\omega)]$ for $\textbf{q}\in\Gamma$-K. The LFE and the SO coupling have sizeable effects on the dielectric screening of bulk Pb, showing an anisotropic behavior. For \textbf{q} vectors in the second BZ, the impact of the LFE on the energy-loss function is remarkable. Inclusion of exchange-correlation effects through the TDLDA kernel increases the intensity of the energy-loss function in the studied range, however without affecting its shape in a significant way.

Very-low energy modes with acoustic-like dispersions are found in all three studied high-symmetry directions and are shown to be a consequence of band structure effects. The character of these acoustic modes depends on the direction of \textbf{q}. The experimental detection of these acoustic modes by electron energy-loss measurements seems feasible as these modes keep their character up to $\omega\simeq$2 eV, and can reach energies as high as 5 eV in the $\Gamma$-X direction.

\section*{ACKNOWLEDGEMENTS}

We acknowledge financial support from the Spanish MICINN (No. FIS2010-19609-C02-01), the Departamento de
Educaci\'on del Gobierno Vasco, and the University of the Basque Country (No. GIC07-IT-366-07).


\begin{thebibliography}{99}

\bibitem{ChulkovCR06} E. V. Chulkov, A. G. Borisov, J. P. Gauyacq, D. S\'anchez-Portal, V.
M. Silkin, V. P. Zhukov, and P. M Echenique, Chem. Rev. (Washington,
D.C.) {\bf 106}, 4160 (2006).

\bibitem{ecbessr04} P. M. Echenique, R. Berndt, E. V. Chulkov, Th. Fauster, A. Goldmann, and U. H\"ofer, Surf. Sci. Rep. {\bf 52}, 219 (2004).

\bibitem{gonzebi} X. Gonze, J.-P. Michenaud, and J.-P. Vigneron, Phys. Rev. B {\bf 41}, 11827 (1990).

\bibitem{zubizarreta} X. Zubizarreta, V. M. Silkin, and E. V. Chulkov, Phys. Rev. B {\bf 84}, 115144 (2011).

\bibitem{Eremeev} S. V. Eremeev, I. A. Nechaev, Yu. M. Koroteev, P. M. Echenique, and E. V. Chulkov, Phys. Rev. Lett. {\bf 108}, 246802 (2012).

\bibitem{heboprb10} R. Heid, K.-P. Bohnen, I. Yu. Sklyadneva, and E. V. Chulkov, Phys. Rev. B {\bf 81} 174527 (2010).

\bibitem{biph} L. E. D\'iaz-S\'anchez, A. H. Romero, and X. Gonze, Phys. Rev. B {\bf 76} 104302 (2007).

\bibitem{Sklyadneva} I. Yu. Sklyadneva, R. Heid, K.-P. Bohnen, V. Chis, V. A. Volodin, K. A. Kokh, O. E. Tereshchenko, P. M. Echenique, and E. V. Chulkov, Phys. Rev. B {\bf 86} 094302 (2012).

\bibitem{glamnjp10} K. Glantschnig and C. Ambrosch-Draxl, New J. Phys. {\bf 12}, 103048 (2010).

\bibitem{dyroprb75} R. C. Dynes and J. M. Rowell, Phys. Rev. B {\bf 11}, 1884 (1975).
%\bibitem{Sklyadneva2011} I. Yu. Sklyadneva, G. Benedek, E. V. Chulkov, P. M. Echenique, R. Heid, K.-P. Bohnen, and J. P. Toennies, Phys. Rev. Lett. %{\bf 107}, 095502 (2011).

\bibitem{pino66} D. Pines  and P. Nozi\`eres, \textit{The Theory of Quantum Liquids} (Benjamin, New York, 1966).

\bibitem{cazaliprb00} M. A. Cazalilla, J. S. Dolado, A. Rubio, and P. M. Echenique, Phys. Rev. B {\bf 61}, 8033 (2000).

\bibitem{aryaseprl94} F. Aryasetiawan and K. Karlsson, Phys. Rev. Lett. {\bf 73}, 1679 (1994).

\bibitem{zhsiprb01} V. P. Zhulkov, V. M. Silkin, E. V. Chulkov, and
P. M. Echenique, Phys. Rev. B {\bf 64}, 180507(R) (2001).

\bibitem{kupiprl02} W. Ku, W. E. Pickett, R. T. Scalettar, and A. G.
Eguiluz, Phys. Rev. Lett. {\bf 88}, 057001 (2002).

\bibitem{ecchprb12} J. P. Echeverry, E. V. Chulkov, P. M. Echenique,
and V. M. Silkin, Phys. Rev. B {\bf 85}, 205135 (2012).

\bibitem{pines56} D. Pines, Can. J. Phys. {\bf 34}, 1379 (1956).

\bibitem{pinespr58} P. Nozi\`eres and D. Pines, Phys. Rev. {\bf 109}, 1062 (1958).

\bibitem{ishiiprb93} Y. Ishii and J. Ruvalds, Phys. Rev. B {\bf 48}, 3455 (1993).

\bibitem{euro} V. M. Silkin, A. Garc\'ia-Lekue, J. M. Pitarke, E. V. Chulkov, E. Zaremba, and P. M. Echenique, Europhys. Lett. {\bf 66}, 260 (2004).

\bibitem{Silkin05} V. M. Silkin, J. M. Pitarke, E. V. Chulkov, and P. M. Echenique, Phys. Rev. B {\bf 72}, 115435 (2005).

\bibitem{nature} B. Diaconescu, K. Pohl, L. Vattuone, L. Savio, P. Hofmann, V. M. Silkin, J. M. Pitarke, E. V. Chulkov, P. M. Echenique, D. Far\'{\i}as, and M. Rocca, Nature (London) {\bf 448}, 57 (2007).

\bibitem{papaprl10} S. J. Park and E. E. Palmer, Phys. Rev. Lett. {\bf 105} 016801 (2010).

\bibitem{podiepl10} K. Pohl, B. Diaconescu, G. Vercelli, L. Vattuone, V. M. Silkin, E. V. Chulkov, P. M. Echenique, and M. Rocca, EPL {\bf 90}, 57006 (2010).

\bibitem{silkinprb09a} V. M. Silkin, A. Balassis, P. M. Echenique, and E. V. Chulkov, Phys. Rev. B {\bf 80}, 054521 (2009).

\bibitem{balassprb08} A. Balassis, E. V. Chulkov, P. M. Echenique, and V. M. Silkin, Phys. Rev. B {\bf 78}, 224502 (2008).

\bibitem{silkinprb09b} V. M. Silkin, I. P. Chernov, Yu. M. Koroteev, and E. V. Chulkov, Phys. Rev. B {\bf 80}, 245114 (2009).

\bibitem{faarprb12} M. Faraggi, A. Arnau, and V. M. Silkin, Phys. Rev. B {\bf 86}, 035115 (2012).

\bibitem{cugaprb12} P. Cudazzo, M. Gatti, and A. Rubio, Phys. Rev. B {\bf 86}, 075121 (2012).

\bibitem{rungeprl84} E. Runge and E. K. U. Gross, Phys. Rev. Lett. {\bf 52}, 997 (1984).

\bibitem{petersprl96} M. Petersilka, U. J. Gossmann, and E. K. U. Gross, Phys. Rev. Lett. {\bf 76}, 1212 (1996).

\bibitem{gross96} E. K. U. Gross, J. F. Dobson, and M. Petersilka, in \textit{Density Functional Theory II}, edited by R. F. Nalewajski (Springer, Berlin, 1996).

\bibitem{aryaseprb94} F. Aryasetiawan and O. Gunnarsson, Phys. Rev. B {\bf 49}, 16214 (1994).

\bibitem{aryase01} F. Aryasetiawan, in \textit{Strong Coulomb Correlations in Electronic Structure Calculations},
edited by V. I. Anisimov (Gordon and Beach, Singapore, 2001).

\bibitem{adpr62} S. L. Adler, Phys. Rev. {\bf 126}, 413 (1962).

\bibitem{bacheleprb82} G. B. Bachelet, D. R. Hamann, and M. Schl\"uter, Phys. Rev. B {\bf 26}, 4199 (1982).

\bibitem{pezuprb81}  J. P. Perdew and A. Zunger, Phys. Rev. B {\bf 23}, 5048 (1981).

\bibitem{cealprl80}  D. M. Ceperley and B. J. Alder, Phys. Rev. Lett. {\bf 45}, 566 (1980).

\bibitem{mp} H. J. Monkhorst and J. D. Pack, Phys. Rev. B {\bf 13}, 5188 (1976).

\bibitem{tinkam71}  M. Tinkam, \textit{Group Theory and Quantum Mechanics} (McGraw-Hill, New York, 1971).

\bibitem{vetoprb08}  M. J. Verstraete,  M. Torrent, F. Jollet, G. Z\'erah, and X. Gonze, Phys. Rev. B {\bf 78}, 045119 (2008).

\bibitem{jepoprb90}  G. J\'ez\'equel and I. Pollini, Phys. Rev. B {\bf 41}, 1327 (1990).

%\bibitem{chkoss91} E. V. Chulkov, Y. M. Koroteev, and V. M. Silkin, Surf. Sci. {\bf 247}, 115 (1991).

\bibitem{ashtonjpf73} A. M. Ashton and G. W. Green, J. Phys. F: Metal Phys. {\ 3}, 179 (1973).

\bibitem{mathewps71} A. G. Mathewson and H. P. Myers, Phys. Scr. {\ 4}, 291 (1971).

\bibitem{wegljpcrd09} W. S. M. Werner, K. Glantschnig, and C. Ambrosch-Draxl, J. Phys. Chem. Ref. Data {\bf 38}, 1013 (2009).

\bibitem{leprprb73} J. C. Lemonnier, M. Priol, and S. Robin, Phys. Rev. B {\bf 8}, 5452 (1973).

\bibitem{erroprb10} I. Errea, A. Rodriguez-Prieto, B. Rousseau, V. M. Silkin, and A. Bergara, Phys. Rev. B {\bf 81}, 205105 (2010).

\bibitem{grosso} G. Grosso and G. P. Parravicini, \textit{Solid State Physics} (Academic Press, San Diego, 2000).

\bibitem{ashcroft} N. W. Ashcroft and N. D. Mermin, \textit{Solid State Physics} (Thomson Learning, Southbank, Victoria, 1976).

\bibitem{Bardeen1961} J. Bardeen and J. R. Schrieffer, in \textit{Progress in Low Temperature Physics}, edited by C. J. Groter (Interscience Publishers, Inc., New York, 1961).

\bibitem{hobrprb09}  I-P. Hong, C. Brun, F. Patthey, I. Yu. Sklyadneva, X. Zubizarreta, R. Heid,
V. M. Silkin, P. M. Echenique, K. P. Bohnen, E. V. Chulkov, and W.-D. Schneider, Phys. Rev. B {\bf 80}, 081409(R) (2009).

\bibitem{Kuzmany1998} H. Kuzmany, \textit{Solid-State Spectroscopy. An introduction} (Springer, Berlin Heidelberg, 1998).

\end{thebibliography}
\end{document}